\documentstyle[twoside,epsf,latexsym,amsmath,amssymb,amsfonts,bbm]{article}

\catcode`\@=11
\long\def\@makefntext#1{
\protect\noindent \hbox to 3.2pt {\hskip-.9pt  
$^{{\eightrm\@thefnmark}}$\hfil}#1\hfill}               %CAN BE USED 

\def\thefootnote{\fnsymbol{footnote}}
\def\@makefnmark{\hbox to 0pt{$^{\@thefnmark}$\hss}}    %ORIGINAL 
        
\def\ps@myheadings{\let\@mkboth\@gobbletwo
\def\@oddhead{\hbox{}
\rightmark\hfil\eightrm\thepage}   
\def\@oddfoot{}\def\@evenhead{\eightrm\thepage\hfil
\leftmark\hbox{}}\def\@evenfoot{}
\def\sectionmark##1{}\def\subsectionmark##1{}}

\oddsidemargin=\evensidemargin
\renewcommand{\thefootnote}{\fnsymbol{footnote}}

\newcounter{sectionc}\newcounter{subsectionc}\newcounter{subsubsectionc}
\renewcommand{\section}[1] {\vspace{12pt}\addtocounter{sectionc}{1} 
\setcounter{subsectionc}{0}\setcounter{subsubsectionc}{0}\noindent 
        {\tenbf\thesectionc. #1}\par\vspace{5pt}}
\renewcommand{\subsection}[1] {\vspace{12pt}\addtocounter{subsectionc}{1} 
\setcounter{subsubsectionc}{0}\noindent 
{\bf\thesectionc.\thesubsectionc. {\kern1pt \bfit #1}}\par\vspace{5pt}}
\renewcommand{\subsubsection}[1] {\vspace{12pt}\addtocounter{subsubsectionc}{1}
        \noindent{\tenrm\thesectionc.\thesubsectionc.\thesubsubsectionc.
        {\kern1pt \tenit #1}}\par\vspace{5pt}}
\newcommand{\nonumsection}[1] {\vspace{12pt}\noindent{\tenbf #1}
        \par\vspace{5pt}}

\newcounter{appendixc}
\newcounter{subappendixc}[appendixc]
\newcounter{subsubappendixc}[subappendixc]
\renewcommand{\thesubappendixc}{\Alph{appendixc}.\arabic{subappendixc}}
\renewcommand{\thesubsubappendixc}
        {\Alph{appendixc}.\arabic{subappendixc}.\arabic{subsubappendixc}}

\renewcommand{\appendix}[1]{\vspace{12pt}
        \refstepcounter{appendixc}
        \setcounter{figure}{0}
        \setcounter{table}{0}
        \setcounter{lemma}{0}
        \setcounter{theorem}{0}
        \setcounter{corollary}{0}
        \setcounter{definition}{0}
        \setcounter{equation}{0}
        \renewcommand{\thefigure}{\Alph{appendixc}.\arabic{figure}}
        \renewcommand{\thetable}{\Alph{appendixc}.\arabic{table}}
        \renewcommand{\theappendixc}{\Alph{appendixc}}
        \renewcommand{\thelemma}{\Alph{appendixc}.\arabic{lemma}}
        \renewcommand{\thetheorem}{\Alph{appendixc}.\arabic{theorem}}
        \renewcommand{\thedefinition}{\Alph{appendixc}.\arabic{definition}}
        \renewcommand{\thecorollary}{\Alph{appendixc}.\arabic{corollary}}
        \renewcommand{\theequation}{\Alph{appendixc}.\arabic{equation}}
        \noindent{\tenbf Appendix \theappendixc #1}\par\vspace{5pt}}
\newcommand{\subappendix}[1] {\vspace{12pt}
        \refstepcounter{subappendixc}
        \noindent{\bf Appendix \thesubappendixc. {\kern1pt \bfit #1}}
        \par\vspace{5pt}}
\newcommand{\subsubappendix}[1] {\vspace{12pt}
        \refstepcounter{subsubappendixc}
        \noindent{\rm Appendix \thesubsubappendixc. {\kern1pt \tenit #1}}
        \par\vspace{5pt}}

\newcommand{\textlineskip}{\baselineskip=13pt}
\newcommand{\smalllineskip}{\baselineskip=10pt}

\newcommand{\copyrightheading}[1]
        {\vspace*{-2.5cm}\smalllineskip{\flushleft
        {\footnotesize Quantum Information and Computation, Vol.~1, No.~0 (2001) 000--000 #1}\\
        {\footnotesize \copyright\kern2pt Rinton Press}\\
         }}

\newcommand{\publisher}[2]{{\begin{center}\footnotesize\smalllineskip 
        Received #1\\
        Revised #2
        \end{center}
        }}

\def\abstracts#1#2#3{{
        \centering{\begin{minipage}{4.5in}\footnotesize\baselineskip=10pt
        \parindent=0pt #1\par 
        \parindent=15pt #2\par
        \parindent=15pt #3
        \end{minipage}}\par}} 

\def\keywords#1{{
        \centering{\begin{minipage}{4.5in}\footnotesize\baselineskip=10pt
        {\footnotesize\it Keywords}\/: #1
         \end{minipage}}\par}}
\def\communicate#1{{
        \centering{\begin{minipage}{4.5in}\footnotesize\baselineskip=10pt
        {\footnotesize\it Communicated by}\/: #1
         \end{minipage}}\par}}

\renewenvironment{thebibliography}[1]
        {\frenchspacing
         \ninerm\baselineskip=11pt
         \begin{list}{\arabic{enumi}.}
        {\usecounter{enumi}\setlength{\parsep}{0pt}     
         \setlength{\leftmargin 12.7pt}{\rightmargin 0pt}%FOR 1--9 ITEMS
         \setlength{\itemsep}{0pt} \settowidth
        {\labelwidth}{#1.}\sloppy}}{\end{list}}

\newcounter{itemlistc}
\newcounter{romanlistc}
\newcounter{alphlistc}
\newcounter{arabiclistc}

\newcommand{\fcaption}[1]{
        \refstepcounter{figure}
        \setbox\@tempboxa = \hbox{\footnotesize Fig.~\thefigure. #1}
        \ifdim \wd\@tempboxa > 5in
           {\begin{center}
        \parbox{5in}{\footnotesize\smalllineskip Fig.~\thefigure. #1}
            \end{center}}
        \else
             {\begin{center}
             {\footnotesize Fig.~\thefigure. #1}
              \end{center}}
        \fi}

\newcommand{\tcaption}[1]{
        \refstepcounter{table}
        \setbox\@tempboxa = \hbox{\footnotesize Table~\thetable. #1}
        \ifdim \wd\@tempboxa > 5in
           {\begin{center}
        \parbox{5in}{\footnotesize\smalllineskip Table~\thetable. #1}
            \end{center}}
        \else
             {\begin{center}
             {\footnotesize Table~\thetable. #1}
              \end{center}}
        \fi}

\def\pmb#1{\setbox0=\hbox{#1}
        \kern-.025em\copy0\kern-\wd0
        \kern.05em\copy0\kern-\wd0
        \kern-.025em\raise.0433em\box0}

\def\fnt#1#2{\footnotetext{\kern-.3em
        {$^{\mbox{\scriptsize #1}}$}{#2}}}

\def\fpage#1{\begingroup
\voffset=.3in
\thispagestyle{empty}\begin{table}[b]\centerline{\footnotesize #1}
        \end{table}\endgroup}

\def\runninghead#1#2{\pagestyle{myheadings}
\markboth{{\protect\footnotesize\it{\quad #1}}\hfill}
{\hfill{\protect\footnotesize\it{#2\quad}}}}
\font\tenrm=cmr10
\font\tenit=cmti10 
\font\tenbf=cmbx10
\font\bfit=cmbxti10 at 10pt
\font\ninerm=cmr9

\font\eightrm=cmr8

\def\FigName{figure}%
\newbox\captionbox
\long\def\@makecaption#1#2{%
  \ifx\FigName\@captype
    \vskip\abovecaptionskip
    \setbox\tempbox\hbox{{\figurecaptionfont #1\hskip1em #2}}
        \ifdim\wd\tempbox< 28pc
        \centerline{\box\tempbox}
        \else
        {\figurecaptionfont #1\hskip1em #2\par}
\fi\else
        \setbox\tempbox\hbox{{\tablecaptionfont #1\hskip1em #2}}
        \ifdim\wd\tempbox< 28pc 
        \centerline{\box\tempbox}
        \else
        {\tablecaptionfont #1\hskip1em #2\par}%
        \fi   
 \vskip\belowcaptionskip
 \fi}
\InputIfFileExists{psfig.sty}
{\typeout{^^Jpsfig.sty inputed...ok}}{\typeout{^^JWarning: psfig.sty could be be found.^^J}}
\InputIfFileExists{epsfsafe.tex}
{\typeout{^^Jepsfsafe.tex inputed...ok}}
                        {\typeout{^^JWarning: epsfsafe.tex could not be found.^^J}}
\InputIfFileExists{epsfig.sty}
{\typeout{^^Jepsfig.sty inputed...ok}}{\typeout{^^JWarning: epsfig.sty could not be found.^^J}}
\InputIfFileExists{epsf.sty}
{\typeout{^^Jepsf.sty inputed...ok}}{\typeout{^^JWarning: epsf.sty could not be found.^^J}}%
\def\fps@figure{tbp}
\def\ftype@figure{1}
\def\ext@figure{lof}
\def\fnum@figure{Fig.\ \thefigure}
\def\qed{\hbox{${\vcenter{\vbox{                  %HOLLOW SQUARE
   \hrule height 0.4pt\hbox{\vrule width 0.4pt height 6pt
   \kern5pt\vrule width 0.4pt}\hrule height 0.4pt}}}$}}

\renewcommand{\thefootnote}{\fnsymbol{footnote}}  %USE SYMBOLIC FOOTNOTE

\def\Tr{{\rm Tr}}
\def\rank{{\rm rank}}
\def\Rea#1{{\rm Re}\left(#1\right)}
\def\Ima#1{{\rm Im}\left(#1\right)}
\def\id{{\mathbbm{1}}}
\def\bra#1{\langle#1|}
\def\ket#1{|#1\rangle}
\def\braket#1#2{\langle#1|#2\rangle}
\def\mean#1{\left<#1\right>}
\def\brac#1{\left(#1\right)}
\def\abs#1{\left|#1\right|}
\def\Cass{{\cal C}^{\sharp}}
\def\Clocass{{\cal C}^{\flat}}
\def\ClocassPOVM{{\cal C}^{\natural}}      
\def\Eass{E^{\sharp}}
\def\Elocass{E^{\natural}}
\begin{document}
\setlength{\textheight}{8.0truein}    %FOR 2ND PAGE ONWARDS

\runninghead{Local vs. joint measurements for the entanglement of assistance} 
            {T. Laustsen, F. Verstraete, and S. J. van Enk}

\normalsize\textlineskip
\thispagestyle{empty}
\setcounter{page}{1}

\copyrightheading{}     %       {Vol.~1, No.~0 (2001) 000--000}

\vspace*{0.88truein}

\fpage{1}
\centerline{\bf LOCAL VS. JOINT MEASUREMENTS}
\vspace*{0.035truein}
\centerline{\bf FOR THE ENTANGLEMENT OF ASSISTANCE}
\vspace*{0.37truein}
\centerline{\footnotesize 
%%%%%%%%%%%%%%%%%%%%%%%%%%%%%%%%%%%%
%put authors' name and address here
%%%%%%%%%%%%%%%%%%%%%%%%%%%%%%%%%%%%
T. LAUSTSEN,${}^{1,2}$ F. VERSTRAETE,${}^{1,3}$ and S. J. VAN ENK${}^{1}$ 
%\footnote{Typeset names in
%10 pt Times Roman, uppercase. Use the footnote to indicate the
%present or permanent address of the author.}
}
\vspace*{0.015truein}
\centerline{\footnotesize\it 
  ${}^{1}$Bell Labs, Lucent Technologies,
  600-700 Mountain Ave,
  Murray Hill, NJ 07974, USA}
\centerline{\footnotesize\it 
  ${}^{2}$QUANTOP -- Danish Quantum Optics Center,
  Department of Physics and Astronomy,}
\baselineskip=10pt
\centerline{\footnotesize\it 
  University of Aarhus,
  DK-8000 \AA rhus C, Denmark}
\centerline{\footnotesize\it 
  ${}^{3}$Department of Mathematical Physics and Astronomy,} 
\baselineskip=10pt
\centerline{\footnotesize\it 
  Ghent University,
  Krijgslaan 281 (S9), 
  B-9000 Gent, Belgium}
\vspace*{0.225truein}
\publisher{(received date)}{(revised date)}

\vspace*{0.21truein}
\abstracts{
We consider a variant of the entanglement of assistance, as independently introduced by D.P.~DiVincenzo {\em et al.} ({\tt quant-ph/9803033}) and O.~Cohen (Phys. Rev. Lett. {\bf 80}, 2493 (1998)). Instead of considering three-party states in which one of the parties, the assistant, performs a measurement such that the remaining two parties are left with on average as much entanglement as possible, we consider four-party states where two parties play the role of assistants. We answer several questions that arise naturally in this scenario, such as (i) how much more entanglement can be produced when the assistants are allowed to perform joint measurements, (ii) for what type of states are local measurements sufficient, (iii) is it necessary for the second assistant to know the measurement outcome of the first, and (iv) are projective measurements sufficient or are more general POVMs needed?}{}{}

\vspace*{10pt}
\vspace*{3pt}
\communicate{to be filled by the Editorial}

\vspace*{1pt}\textlineskip      %) USE THIS MEASUREMENT WHEN THERE IS
\section{Introduction}          %) A SECTION HEADING
\vspace*{-0.5pt}
\label{sec:introduction}
\setcounter{footnote}{0}
\renewcommand{\thefootnote}{\alph{footnote}}
\noindent Consider a pure state of four qubits located at parties A, B, C, and D. Whether this state is entangled or not, the reduced density matrix of the AB system, $\rho_{{\rm AB}}$, might only possess very little entanglement, or it might even be separable. But the parties C and D, both knowing the full state, can help increasing this entanglement by performing measurements on their qubits and communicating their results to A and B. This leads to the concept of the \emph{entanglement of assistance} \cite{art211}, defined as the maximum possible average entanglement the assistant parties A and B can create between C and D. This was originally introduced by Cohen \cite{art292} in the notion of \emph{maximal hidden entanglement} and later independently rediscovered by the authors of Ref.~\cite{art211}. It follows from a result of Hughston, Jozsa, and Wootters (HJW) \cite{art129} that if the assistant parties are allowed to perform joint measurements, then the entanglement of assistance depends only on $\rho_{{\rm AB}}$, and as such it is considered in Ref.~\cite{art211} as a measure of two-party entanglement. We denote this quantity by $\Eass$\footnote{We use a musical notation where $\natural$ (`natural') signifies that only local measurements (and classical communication) are included, whereas $\sharp$ (`sharp') signifies that joint measurements are included. Later on we will also use $\flat$ (`flat') to signify that only local projective measurements are included.}. Here we ask the question, what is the entanglement of assistance if the assistant parties are restricted to performing only local measurements? This quantity, which we denote $\Elocass$, is not intended as a new measure of two-party entanglement, indeed it will in general depend on the full four-party state. Clearly $\Elocass\le \Eass$, but can this upper bound be reached, and if so for what states? 

\begin{figure} [htbp]
\vspace*{13pt}
\centerline{\psfig{file=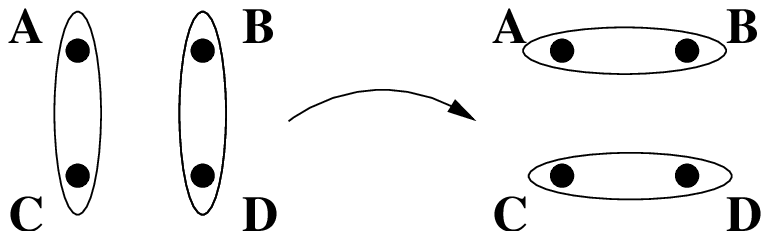, width=8.2cm}} %100 percent
\vspace*{13pt}
\fcaption{\label{fig:swap}
A simple example of a state where joint measurements can create entanglement across a separable cut where local measurements cannot, is the entanglement swapping indicated here. C and D make a joint Bell state measurement on the initial state, Eq.~(\ref{eq:8}), thereby creating a maximally entangled state between A and B. For this state the entanglement of local assistance is zero.}
\end{figure}

Before attacking the problem we look at some simple examples to catch a glimpse of the properties of $\Elocass$ in relation to $\Eass$. For example consider the four-party GHZ state,
\noindent
\begin{equation}
\label{eq:7}
\ket{\psi}_{{\rm ABCD}}=\frac{\ket{0000}+\ket{1111}}{\sqrt{2}}\,.
\end{equation}
If C and D both measure in the basis $\{\ket{+},\ket{-}\}$, defined as $\ket{\pm}=\frac{1}{\sqrt{2}}\brac{\ket{0}\pm\ket{1}}$, A and B will always end up with a maximally entangled state, so that in this case the two quantities are equal, $\Eass=\Elocass=1$\footnote{Here we use the standard measure of entanglement, $E(\ket{\psi})=-\Tr\rho_A\log_2\rho_A$, see Ref.~\cite{art41}.}. On the other hand, for the state\footnote{Here and throughout we have denoted the four maximally entangled Bell states $\ket{\Phi^{\pm}}=\frac{1}{\sqrt{2}}\brac{\ket{00}\pm\ket{11}}$ and $\ket{\Psi^{\pm}}=\frac{1}{\sqrt{2}}\brac{\ket{01}\pm\ket{10}}$.}
\noindent
\begin{equation}
\label{eq:8}
\ket{\psi}_{{\rm ABCD}}=\ket{\Phi^{+}}_{{\rm AC}}\ket{\Phi^{+}}_{{\rm BD}}\,,
\end{equation}
no local measurements by C and D can change the fact that A and B are separable, and hence $\Elocass=0$. But with a joint measurement in the Bell state basis, C and D can ensure that A and B end up in a maximally entangled state, as illustrated in Fig.~\ref{fig:swap}, so that $\Eass=1$ for this state. This is obviously the largest difference we can get between $\Eass$ and $\Elocass$.

A result which is useful for a certain class of states is the one by Walgate \emph{et al.} \cite{art402}, stating that two orthogonal states of two qubits can always be locally distinguished. This result can be applied to immediately determine a set of states for which local measurements are sufficient. Indeed for the states where the optimal joint measurement is a projection onto at most two orthogonal states (with nonzero probability) we have $\Elocass=\Eass$. A particular local measurement that achieves this can be found in Ref.~\cite{art402}. For this measurement we note that classical communication of the outcome of C's measurement to D is explicitly needed.

Yet another example of a state for which communication between the assistants seems to be necessary is
\noindent
\begin{equation}
\label{eq:75}
\ket{\psi}_{{\rm ABCD}}=
\frac{1}{4}\Big(\ket{\Phi^{+}}_{{\rm AB}}\ket{00}_{{\rm CD}}
                +\ket{\Phi^{-}}_{{\rm AB}}\ket{01}_{{\rm CD}}
               +\ket{\Psi^{+}}_{{\rm AB}}\ket{1+}_{{\rm CD}}
                +\ket{\Psi^{-}}_{{\rm AB}}\ket{1-}_{{\rm CD}}\Big)\,.
\end{equation}
For this state one optimal local measurement is clearly the one where C measures in the $\{\ket{0},\ket{1}\}$-basis and D measures either in the $\{\ket{0},\ket{1}\}$-basis or in the $\{\ket{+},\ket{-}\}$-basis depending on the outcome of C's measurement. However, we will show in Sec.~\ref{sec:class-comm-not} that, rather surprisingly, classical communication of measurement outcomes between the assistant parties is never necessary, as long as they restrict themselves to von Neumann measurements. In particular we note that for the above state, Eq.~(\ref{eq:75}), C can measure in the $\{\ket{0},\ket{1}\}$-basis and D in the $\{\frac{\ket{0}+i\ket{1}}{\sqrt{2}},\frac{\ket{0}-i\ket{1}}{\sqrt{2}}\}$-basis irrespective of C's measurement outcome. 

We note that it might be necessary for C and D to communicate to decide on which optimal basis to use, if there is no unique such basis. Moreover classical communication from CD to AB will of course always be necessary. Otherwise, from their point of view, AB would be stuck in the state $\rho_{{\rm AB}}$ no matter what is happening at C and D.

For convenience we will take the concurrence of Wootters \cite{art280,art83} as our measure of entanglement. Concurrence is a well-established measure of entanglement, in particular it is an entanglement monotone \cite{art82}. Moreover, the concurrence features in some elegant inequalities \cite{art148}, see also \cite{art293}, that are not satisfied by the entanglement of formation.

Thus we introduce the \emph{concurrence of assistance}. When we allow joint measurements we will denote this quantity by $\Cass$. When we allow only local measurements we will distinguish two quantities, namely $\Clocass$ where we allow only local von Neumann measurements, and $\ClocassPOVM$ where we allow more general POVMs. In the main part of the present paper we will restrict the assistant parties to local von Neumann measurements and classical communication. This as well as a scrutiny of how much freedom we have in choosing optimal (local or joint) measurements for a given state, will be the topic of Sec.~\ref{sec:conc-assist}. In Sec.~\ref{sec:class-comm-not} we will show that classical communication of measurement outcomes between the assistants is not necessary. In Sec.~\ref{sec:what-states-are} we will find a decomposition that allows us to quickly determine (i) the concurrence of assistance, (ii) whether there is a local measurement with which we can achieve this, as well as (iii) the explicit form of this local measurement. In Sec.~\ref{sec:numerical-results} we will present some numerical results showing the advantage of joint measurements on arbitrary states, and in Sec.~\ref{sec:discussion} we discuss the effect of non-projective measurements and some possible extensions of the present work.

\section{Concurrence of Assistance}
\label{sec:conc-assist}
\noindent The entanglement of assistance has been considered by DiVincenzo \emph{et al.}, Ref.~\cite{art211}. Starting from a three party pure state $\ket{\psi}_{{\rm ABC}}$, it quantifies how much pure state entanglement party C can make between A and B by doing local measurements on his part of the state. 

The reduced state $\rho_{{\rm AB}}$ can be described by infinitely many different pure state ensembles, ${\cal E}=\{ p_i,\ket{\phi_{i}}\}$ all satisfying $\rho_{{\rm AB}}=\sum_i p_i \ket{\phi_i}\bra{\phi_i}$, and from each ensemble we can construct a purification as 
\noindent
\begin{equation}
\label{eq:69}
\ket{\psi_{{\cal E}}}=\sum_i \sqrt{p_i}\ket{\phi_i}_{{\rm AB}}\otimes\ket{i}_{{\rm C}}\,,
\end{equation}
where $\{\ket{i}_{{\rm C}}\}$ is an orthonormal basis of C's system, which might have been extended by an ancilla system to increase its dimension. Only one of these purifications is the state $\ket{\psi}_{{\rm ABC}}$, but the HJW-theorem \cite{art129}, mentioned above, states that we can transform purifications corresponding to different ensembles into each other by a local unitary operation acting on the system of the assistant party C alone, which means that, in particular, all the states $\ket{\psi_{{\cal E}}}$ of Eq.~(\ref{eq:69}) can be reached from our state $\ket{\psi}_{{\rm ABC}}$ by local operations on this system.

A local measurement by C in the basis $\{\ket{i}_{{\rm C}}\}$ will make the AB-system end up in the state $\ket{\phi_i}$ with probability $p_i$, thus realizing ${\cal E}$ as a classical statistical ensemble. Conversely any measurement that C can do on the state, will correspond to such an ensemble. These considerations led the authors of Ref.~\cite{art211} to define the entanglement of assistance as
\noindent
\begin{equation}
\label{eq:1}
\Eass=\max_{{\cal E}}\sum_i p_i E\brac{\ket{\phi_{i}}}\,.
\end{equation}
The measure of entanglement, $E(\ket{\phi_{i}})$, used here is the von Neumann entropy of the marginal density matrix $\rho_{{\rm A}}=\Tr_{{\rm B}}\ket{\phi_i}\bra{\phi_i}$, coinciding for pure states with such other measures as the entanglement of formation and the entanglement of distillation, see e.g. Ref.~\cite{art41}. 

We see that the entanglement of assistance is a function of the reduced state $\rho_{{\rm AB}}$ only, and as such it is suggested in \cite{art211} as a measure of two-party entanglement. An explicit formula has not been found, but for the case where the systems of A and B are qubits, it was shown in \cite{art211} that $\Eass\le F(\rho,\tilde{\rho})$, where $\rho=\rho_{{\rm AB}}$, and where the fidelity function is defined as
\noindent
\begin{equation}
\label{eq:11}
F(\rho,\sigma)=\Tr\sqrt{\rho^{1/2}\sigma\rho^{1/2}}\,.
\end{equation}
Here the Wootters tilde, Refs.~\cite{art83,art280}, is defined for pure states and mixed states respectively as 
\noindent
\begin{align}
\ket{\tilde{\psi}}&=\brac{\sigma_y\otimes\sigma_y}\ket{\psi^{*}}\,,\nonumber\\
\tilde{\rho}&=\brac{\sigma_y\otimes\sigma_y}\rho^{*}\brac{\sigma_y\otimes\sigma_y}\,,\label{eq:6}
\end{align}
with the complex conjugation taken in the standard basis. 

In these two references, see also Ref.~\cite{art403}, an alternative measure of entanglement for two qubits, namely the concurrence, is defined. For pure states $\ket{\psi}=a\ket{00}+b\ket{01}+c\ket{10}+d\ket{11}$ this is 
\noindent
\begin{equation}
\label{eq:70}
C\brac{\ket{\psi}}=|\braket{\psi}{\tilde{\psi}}|=2\abs{ad-bc}\,.
\end{equation}
The main advantage of this measure is its simple algebraic form, which enables an analytic treatment of some of the problems to be presented in this paper. In analogy to Eq.~(\ref{eq:1}) we therefore define the \emph{concurrence of assistance}\footnote{We remark that the function $E(C)$ relating the concurrence to the entanglement measure used in Eq.~(\ref{eq:1}), is not linear \cite{art83}, so in particular $\Eass\ne E\brac{\Cass}$, and hence the concurrence of assistance is really an alternative measure to the entanglement of assistance, rather than a simple function thereof.}  { }as 
\noindent
\begin{equation}
\label{eq:2}
\Cass=\max_{{\cal E}}\sum_i p_i C\brac{\ket{\phi_{i}}}\,.
\end{equation}
The HJW-theorem ensures also in this case dependence on $\rho_{{\rm AB}}$ only.

One specific ensemble describing our two-qubit state $\rho=\rho_{{\rm AB}}$ is the one found by Wootters in Ref.~\cite{art83}, ${\cal E}_{W}=\{q_i,\ket{x_i}\}$ satisfying $\sqrt{q_iq_j}\braket{x_i}{\tilde{x}_j}=\lambda_i\delta_{ij}$, where $\lambda_i$ denotes the eigenvalues of the matrix $\sqrt{\rho^{1/2}\tilde{\rho}\rho^{1/2}}$. A measurement yielding these $\ket{x_i}$'s with probabilities $q_i$ will result in an average concurrence of
\noindent
\begin{equation}        
\label{eq:3}
\bar{{\cal C}}({\cal E}_W)=\sum_i p_i |\braket{x_i}{\tilde{x}_i}|=\sum_i \lambda_i=F(\rho_{{\rm AB}},\tilde{\rho}_{{\rm AB}})\,.
\end{equation}
The \emph{strong concavity property} of the fidelity function, Theorem 9.7 of Ref.~\cite{NielsenChuang}, tells us that for \emph{any} ensemble ${\cal E}=\{\pi_i,\ket{\phi_i}\}$ describing $\rho_{{\rm AB}}$, this is an upper bound,
\noindent
\begin{align}
\bar{{\cal C}}({\cal E})&=\sum_i \pi_i |\braket{\phi_i}{\tilde{\phi}_i}|
             =\sum_i \pi_i F\brac{\ket{\phi_i}\bra{\phi_i},\ket{\tilde{\phi}_i}\bra{\tilde{\phi}_i}}\nonumber\\
             &\le F\brac{\sum_i \pi_i \ket{\phi_i}\bra{\phi_i},\sum_i \pi_i \ket{\tilde{\phi}_i}\bra{\tilde{\phi}_i}}
             =F(\rho_{{\rm AB}},\tilde{\rho}_{{\rm AB}})\,,\label{eq:4}
\end{align}
so the value (\ref{eq:3}) reached with the ensemble ${\cal E}_W$ is in fact equal to the concurrence of assistance,
\noindent
\begin{equation}
\label{eq:5}
\Cass=F(\rho_{{\rm AB}},\tilde{\rho}_{{\rm AB}})\,,
\end{equation}
as was pointed out by C.A.~Fuchs \cite{note:chris}.

In the following we will focus on pure states of four qubits. For such states a new problem arises, namely the one of \emph{local assistance}. How much concurrence can the assistant parties C and D make for A and B after local measurements on their own qubits, $\ClocassPOVM$, and what measurement is the optimal one? These and other related questions will be the topic of the following sections.

First we derive some general properties of the concurrence of assistance. From now on let
\noindent
\begin{equation}
\label{eq:33}
\ket{\psi}=c_1\ket{0000}+c_2\ket{0001}+\cdots+c_{16}\ket{1111}
\end{equation}
be the initially shared pure state of four qubits located at parties A, B, C, and D respectively, and construct the matrix
\noindent
\begin{equation}
\label{eq:32}
X=\brac{\begin{array}{cccc}
c_1 & c_2 & c_3 & c_4\\
c_5 & c_6 & c_7 & c_8\\
c_9 & c_{10} & c_{11} & c_{12}\\
c_{13} & c_{14} & c_{15} & c_{16}\end{array}}\,.
\end{equation}
By explicitly checking we see that 
\noindent
\begin{equation}
\label{eq:35}
\rho_{{\rm AB}}=XX^{\dagger}\,,
\end{equation}
and by symmetry it then follows that $\rho_{{\rm CD}}=X^T(X^T)^{\dagger}$. A useful way to think of $X$ is that if we write $\ket{\psi}=\ket{\phi_{00}}\ket{00}+\cdots+\ket{\phi_{11}}\ket{11}$, then $X$ is the matrix with the unnormalized $\ket{\phi_{ij}}$'s as column vectors,
\noindent
\begin{equation}
\label{eq:38}
X=\brac{\begin{array}{cccc}
| & | & | & |\\
\ket{\phi_{00}} & \ket{\phi_{01}} & \ket{\phi_{10}} & \ket{\phi_{11}}\\
| & | & | & |\end{array}}\,.
\end{equation}
A measurement done by C and D in their $\{\ket{0},\ket{1}\}$-bases with outcome $\ket{i}_{{\rm C}}\ket{j}_{{\rm D}}$ will project the state of the AB-system onto $\ket{\phi_{ij}}$, and this outcome will occur with probability $\braket{\phi_{ij}}{\phi_{ij}}$. The concurrence of this state, Eq.~(\ref{eq:70}), is $|\braket{\tilde{\phi_{ij}}}{\phi_{ij}}|/\braket{\phi_{ij}}{\phi_{ij}}$, so the average concurrence obtained by this particular measurement is
\noindent
\begin{equation}
\label{eq:39}
\bar{{\cal C}}=\sum_{ij}|\braket{\tilde{\phi_{ij}}}{\phi_{ij}}|\,.
\end{equation}
In terms of the matrix $X$ introduced above this reduces to
\noindent
\begin{equation}
\label{eq:40}
\bar{{\cal C}}
=\sum_{k}\abs{\brac{\brac{\brac{\sigma_y\otimes\sigma_y}X^{*}}^{\dagger}X}_{kk}}
=\sum_{k}\abs{\brac{X^{T}\brac{\sigma_y\otimes\sigma_y}X}_{kk}}\,.
\end{equation}

Transformations of the type 
\noindent
\begin{equation}
\label{eq:21}
X\longrightarrow XV\,,
\end{equation}
where $V$ is a (rectangular) right unitary matrix, $VV^{\dagger}=\id$, do not change $\rho_{{\rm AB}}$ (\ref{eq:35}), and this is indeed the only type of transformation of $X$ that leaves $\rho_{{\rm AB}}$ unchanged. This tells us the freedom we have in choosing an $X$ satisfying Eq.~(\ref{eq:35}) for a given state, which is completely equivalent to the freedom mentioned above (\ref{eq:69}) we have in choosing the ensemble.

Suppose that $V$ is square, and hence unitary (this corresponds to a von Neumann measurement). From Eq.~(\ref{eq:38}) we see that the column vectors of $XV$ are the final states of the AB system after C and D have done a (von Neumann) measurement in the standard basis on the state $(XV)^{T}(XV)^{T\dagger}=V^T\rho_{{\rm CD}}V^{T\dagger}$, or equivalently, a measurement on $\rho_{CD}$ in the basis consisting of the column vectors of $V^{*}$. In general this involves joint actions on the CD system.

The average concurrence of the AB-state after this measurement will be as in Eq.~(\ref{eq:40}), but with $X\to XV$, so the concurrence of assistance is the maximum
\noindent
\begin{equation}
\label{eq:41}
\Cass=\max_V\sum_{k}\abs{\brac{V^{T}QV}_{kk}}\,,
\end{equation}
where the matrix $Q$ is defined as
\noindent
\begin{equation}
\label{eq:42}
Q=X^{T}\brac{\sigma_y\otimes\sigma_y}X\,.
\end{equation}

To find this maximum we apply some mathematical results, the first one being the existence of \emph{Takagi's factorization}, see Ref.~\cite{HornJohnson}.
It states that if $Q$ is a complex symmetric matrix we can find a unitary matrix $U$ such that
\noindent
\begin{equation}
\label{eq:43}
U^{T}QU=\brac{\begin{array}{cccc}
\sigma_1 & 0 & 0 & 0\\
0 & \sigma_2 & 0 & 0\\
0 & 0 & \sigma_3 & 0\\
0 & 0 & 0 & \sigma_4\end{array}}\,,
\end{equation}
where $\sigma_k$ are the singular values of the matrix $Q$.

Our matrix $Q$ of Eq.~(\ref{eq:42}) is indeed symmetric, $Q_{ij}=\sum_{kl}(X^{T})_{ik}(\sigma_y\otimes\sigma_y)_{kl}X_{lj}=\sum_{kl}(X^{T})_{jl}(\sigma_y\otimes\sigma_y)_{lk}X_{ki}=Q_{ji}$, and together with the next mathematical result, namely that
\noindent
\begin{equation}
\label{eq:10}
\sum_k\abs{\brac{V^{T}QV}_{kk}}\le\sum_k\sigma_k\,,
\end{equation}
for all right unitary matrices $V$, see App.~\ref{sec:proof-eq.-refeq:10}, we get the following nice formula for the concurrence of assistance,
\noindent
\begin{equation}
\label{eq:36}
\Cass=\sum_{k}\abs{\brac{U^{T}QU}_{kk}}=\sum_k\sigma_k\,.
\end{equation}
Moreover we get a particular candidate for the optimal measurement, namely the von Neumann measurement projecting onto the column vectors of $U^{*}$, where $U$ is the unitary matrix found from the Takagi decomposition of $Q$, Eq.~(\ref{eq:43}).

This is, however, not the only optimal measurement. For our present problem it is especially interesting to see whether there is an optimal measurement that is local. To see this we need to specify what freedom we have in choosing matrices $V$ that reach the maximum in Eq.~(\ref{eq:41}). For convenience we shall assume that $V$ is unitary, and hence restrict C and D to von Neumann measurements.

Singular values are per definition real and positive, so from equation (\ref{eq:36}), $\Cass=\Tr(U^{T}QU)$ $=\Tr(O^{T}U^{T}QUO)$, showing that the unitary matrix $UO$, where $O$ is an arbitrary real, orthogonal matrix ($O^{T}O=\id$), is another choice of optimal measurement basis. Supposing that all the singular values of $Q$ are nonzero, further multiplication by unitary matrices that does not change the sum of the absolute values in Eq.~(\ref{eq:36}), can only be with a diagonal unitary matrix, since these change only the phases of the entries. So in this case all unitary $V$'s corresponding to an optimal measurement basis are of the form
\noindent
\begin{equation}
\label{eq:37}
V=UO\brac{\begin{array}{cccc}
e^{i\varphi_1} & 0 & 0 & 0\\
0 & e^{i\varphi_2} & 0 & 0\\
0 & 0 & e^{i\varphi_3} & 0\\
0 & 0 & 0 & e^{i\varphi_4}\end{array}}\,,
\end{equation}
where $U$ is a particular matrix satisfying Eq.~(\ref{eq:43}), $O$ is a real, orthogonal matrix and $\varphi_1,\ldots,\varphi_4$ are arbitrary real numbers.

\section{Classical Communication is Not Necessary}
\label{sec:class-comm-not}
\noindent The results of the preceding section do not concern distinguishing local and joint measurements, but we are now in a position to treat this problem. For local measurements we have the possibility of communicating measurement outcomes between the assistant parties, and letting the outcome of one party's measurement determine the measurement basis to be used by the other party. As discussed in the introduction, one might expect this to be necessary for the optimal local measurement, but in this section we shall show that as long as the measurements are von Neumann measurements, classical communication of measurement outcomes between the parties is actually not necessary, no matter what state we are looking at.

To show this, suppose that party C is the first party to do a measurement, and that he does it in the basis $W$ (by ``a measurement in the basis $W$'' we mean ``a measurement in the basis consisting of the column vectors of $W^{*}$'', as discussed above). Party D now has the opportunity to let her measurement basis depend on the outcome of C's measurement, so the unitary matrix $V$ corresponding to their combined measurement will be of the form
\noindent
\begin{equation}
\label{eq:44}
V=(W\otimes\id)\brac{\begin{array}{cc}
V_1 & 0\\
0 & V_2\end{array}}\,,
\end{equation}
where $W$, $V_1$, and $V_2$ are all two-dimensional unitary matrices. We will now show, that among the best measurements that D can do following this measurement by C, there is one satisfying $V_1=V_2$, so that in total $V=W\otimes V_1$, and therefore D doesn't have to wait for the outcome of C's measurement.

The concurrence of assistance of a state, with our restriction to local von Neumann measurements and classical communication, is according to Eq.~(\ref{eq:41})
\noindent
\begin{equation}
\label{eq:45}
\Clocass=\max_{V}\sum_k 
\left|\brac{V^{T}QV}_{kk}\right|\,,
\end{equation}
where the maximum is taken over unitaries, $V$, either of the form (\ref{eq:44}) or of the similar form with the roles of C and D reversed, and where $Q$ is defined by Eq.~(\ref{eq:42}). Let
\noindent
\begin{equation}
\label{eq:46}
\brac{W^{T}\otimes\id}Q\brac{W\otimes\id}=\brac{\begin{array}{cc}Q_1 & *\\ * & Q_2\end{array}}\,,
\end{equation}
where the $*$'s indicate that the off-diagonal blocks are irrelevant. After C's measurement the state has collapsed, so that either the first or the last two columns of the matrix $X(W\otimes\id)$ will be all zeros, see Eq.~(\ref{eq:38}), and consequently the matrix in Eq.~(\ref{eq:46}) will have become one of the two,
\noindent
\begin{equation}
\label{eq:48}
\brac{\begin{array}{cc}Q_1 & 0\\0 & 0\end{array}}
\hspace{.5cm}{\rm or}\hspace{.5cm}
\brac{\begin{array}{cc}0 & 0\\0 & Q_2\end{array}}\,.
\end{equation}
Since $\brac{W^{T}\otimes\id}Q\brac{W\otimes\id}$ is complex symmetric, so are $Q_1$ and $Q_2$, Eq.~(\ref{eq:46}), and we can therefore write the Takagi-decomposition, Eq.~(\ref{eq:43}), of these, $Q_1=\tilde{V}_{1}^{*}\Sigma_1 \tilde{V}_{1}^{\dagger}$ and $Q_2=\tilde{V}_{2}^{*}\Sigma_2 \tilde{V}_{2}^{\dagger}$. From Eqs.~(\ref{eq:44}), ~(\ref{eq:45}), and (\ref{eq:46}) the highest achievable average concurrence following this measurement by C is
\noindent
\begin{equation}
\label{eq:49}
\bar{{\cal C}}_{{\rm max}}=\max_{V_1,V_2}\sum_k 
\left|\brac{\begin{array}{cc}
V_{1}^{T}Q_1V_1 & *\\
* & V_{2}^{T}Q_2V_2\end{array}}_{kk}\right|\,,
\end{equation}
which, by the same argument as in Eq.~(\ref{eq:36}), reduces to
\noindent
\begin{equation}
\label{eq:50}
\bar{{\cal C}}_{{\rm max}}=\Tr\brac{\Sigma_1}+\Tr\brac{\Sigma_2}\,.
\end{equation}
Hence we see that $\tilde{V}_{1}$ and $\tilde{V}_{2}$ are one maximizing choice for $V_1$ and $V_2$, and as in Eq.~(\ref{eq:37}) we see that we will get the same average concurrence by choosing $V_1$ and $V_2$ as
\noindent
\begin{equation}
\label{eq:51}
V_1=\tilde{V}_{1}O_1
\brac{\begin{array}{cc}e^{i\alpha_1} & 0\\0 & e^{i\alpha_2}\end{array}}
\hspace{.5cm}{\rm and}\hspace{.5cm}
V_2=\tilde{V}_{2}O_2
\brac{\begin{array}{cc}e^{i\beta_1} & 0\\0 & e^{i\beta_2}\end{array}}\,,
\end{equation}
where $O_1$, $O_2$ are real orthogonal matrices and $\alpha_{1}$, $\alpha_{2}$, $\beta_1$, and $\beta_2$ are arbitrary real numbers. We want to show that we can find $O_1$ and $O_2$ and a set of phases such that $V_1=V_2$, or equivalently
\noindent
\begin{equation}
\label{eq:52}
\tilde{V}_{1}^{\dagger}\tilde{V}_{2}=O_{1}\brac{\begin{array}{cc}
e^{i(\alpha_1-\beta_1)} & 0\\
0 & e^{i(\alpha_2-\beta_2)}\end{array}}O_{2}^{T}\,.
\end{equation}
But this is in fact always true, due to the following theorem:

Given a unitary $n\times n$ matrix $U$ we can find real orthogonal matrices $O_1$, $O_2$, so that
\noindent
\begin{equation}
\label{eq:53}
U=O_1\brac{\begin{array}{cccc}
e^{i\delta_1} & 0 & \cdots & 0\\
0 & e^{i\delta_2} &   & \vdots\\
\vdots &  & \ddots & 0\\
0 & \cdots & 0 & e^{i\delta_n}
\end{array}}O_{2}^{T}\,.
\end{equation}
The phases, which we will refer to as the \emph{phases of $U$},  $\delta_i$ ($-\frac{\pi}{2}<\delta_i\le\frac{\pi}{2}$) are uniquely defined (up to permutations). The proof of this theorem, as well as how much freedom we have in choosing $O_1$ and $O_2$, is given in Appendix~\ref{sec:proof-eq.-refeq:53}.

Hence it is always possible to fulfil Eq.~(\ref{eq:52}), and therefore for C and D to make an optimal local measurement in a basis of the form $V=W\otimes V_1$. This means that classical communication of measurement outcomes is not necessary to improve the average concurrence achieved in this type of measurements, and in particular we note that it doesn't matter which one of the assistant parties, C and D, performs the measurement first.

Having obtained this result we now also get a simple expression for $\Clocass$. From Eqs.~(\ref{eq:45})~--~(\ref{eq:50}) it immediately follows that
\noindent
\begin{equation}
\label{eq:79}
\Clocass=\max_{W}\brac{\Tr\brac{\Sigma_1}+\Tr\brac{\Sigma_2}}\,, 
\end{equation}
where $W$ describes the measurement done by C, and $\Sigma_1$ and $\Sigma_2$ are the singular value matrices of the two parts of $Q$ after C's measurement, defined in Eq.~(\ref{eq:46}).

\section{For What States are Local Measurements Sufficient?}
\label{sec:what-states-are}
\noindent In this section we are going to find the set of states for which the assistant parties, C and D, have no advantage in using joint measurements as compared to local von Neumann measurements, \emph{i.e.} $\Clocass=\Cass$. Of course this automatically implies that also $\ClocassPOVM=\Cass$.

\subsection{$\rank\brac{\Sigma}=4$}
\label{sec:rankbracsigma=4}
\noindent Our starting point will be the expression, Eq.~(\ref{eq:41}), for the concurrence of assistance. At first we assume that the initial state $\ket{\psi}$ has $\rank\brac{\rho_{{\rm AB}}}=4$, which is equivalent to assuming that all the singular values of the matrix $Q$ of Eq.~(\ref{eq:42}) are nonzero, \emph{i.e.} $\rank\brac{\Sigma}=4$. The remaining cases will be discussed separately below. For the states we consider here, the degree of freedom we have in choosing the optimal measurement basis $V$ is completely determined by equation (\ref{eq:37}). Together with the result of the previous section that classical communication of measurement outcomes is not necessary for this type of measurements, this equation tells us that local von Neumann measurements can do as well as the best joint measurement if and only if we can find matrices $O_1\in SO(4)$, $D\in SU(4)$ (diagonal), and $U_1,U_2\in SU(2)$ such that
\noindent
\begin{equation}
\label{eq:54}
UO_1D=U_1\otimes U_2\,,
\end{equation}
where $U\in SU(4)$ is determined by Eq.~(\ref{eq:36}). For convenience we are requiring all matrices to have determinant one\footnote{This is a legal requirement, since it only amounts to multiplying the matrix with a constant phase factor. This does not change the concurrence, Eq.~(\ref{eq:41}), and only changes the measurement basis (columns of $V^{*}$) by an overall phase factor.}.

Now, define the unitary matrix $T\in SU(4)$ by
\noindent
\begin{equation}
\label{eq:57}
T=\frac{1}{\sqrt{2}}
\brac{\begin{array}{cccc}
1 & 0 & 0 & 1\\
0 & i & i & 0\\
0 & -1 & 1 & 0\\
i & 0 & 0 & -i
\end{array}}\,.
\end{equation}
This matrix has the useful property that for any two matrices $U_1,U_2\in SU(2)$ the matrix $T\brac{U_1\otimes U_2}T^{\dagger}\in SO(4)$, and conversely, for any matrix $O\in SO(4)$ there exist $U_1,U_2\in SU(2)$ such that $T^{\dagger}OT=U_1\otimes U_2$, see Ref.~\cite{art408}.

Applying this result to the criterion Eq.~(\ref{eq:54}), we see that it becomes
\noindent
\begin{equation}
\label{eq:55}
TU=O_2TD^{\dagger}O_{1}^{T}\,,
\end{equation}
where $O_2=T\brac{U_1\otimes U_2}T^{\dagger}$. The matrix $TU$ is unitary so we can apply the decomposition (\ref{eq:53}) to find its phases, \emph{i.e.} to find $O_3,O_4\in SO(4)$ and $E\in SU(4)$ (diagonal) such that 
\noindent
\begin{equation}
\label{eq:63}
TU=O_3EO_{4}^{T}\,.
\end{equation}
The criterion (\ref{eq:55}) now becomes: when can we find $O_5,O_6\in SO(4)$ and $D\in SU(4)$ (diagonal) such that
\noindent
\begin{equation}
\label{eq:56}
E=O_5TD^{\dagger}O_{6}^{T}\,,
\end{equation}
where $E$ is determined from the state $\ket{\psi}$ by Eq.~(\ref{eq:63}). The matrix $TD^{\dagger}$ is also unitary, and accordingly the phases of that matrix can also be determined by writing it in the form of Eq.~(\ref{eq:53}). When varying $D$ there must be some constraints on what the phases of $TD^{\dagger}$ can be. (If all choices were possible, it would mean that local measurements were always sufficient, in contradiction with several counterexamples, as discussed in the introduction).

By explicit calculation, following the lines of App.~\ref{sec:proof-eq.-refeq:53}, we find that with the diagonal entries of $D$ equal to $\{e^{i\varphi_1},e^{i\varphi_2},e^{i\varphi_3},e^{i\varphi_4}\}$, the phases of $TD^{\dagger}$ turn out to be
\noindent
\begin{equation}
\label{eq:62}
\left\{e^{i\Phi},e^{-i\Phi},e^{-i\brac{\Phi-\frac{\pi}{2}}},e^{i\brac{\Phi-\frac{\pi}{2}}}\right\}\,,
\end{equation}
where $\Phi=\brac{\frac{1}{4}(\varphi_1+\varphi_4-\varphi_2-\varphi_3)\textrm{ mod }\frac{\pi}{2}}$ lies in the interval $0<\Phi\le\frac{\pi}{2}$. Hence our criterion for local von Neumann measurements to be sufficient is that there has to exist a number $\Phi$ such that the phases of $TU$, Eq.~(\ref{eq:63}), are of the form (\ref{eq:62}). 

$U$ is found from the state $X$ according to (\ref{eq:42}),~(\ref{eq:43}),
\noindent
\begin{equation}
\label{eq:64}
U^{T}X^{T}\brac{\sigma_y\otimes\sigma_y}XU=\Sigma\,,
\end{equation}
where $\Sigma$ is a diagonal matrix containing the singular values of $Q=X^{T}\brac{\sigma_y\otimes\sigma_y}X$, which according to Eq.~(\ref{eq:36}) gives us the concurrence of assistance,
\noindent
\begin{equation}
\label{eq:12}
\Cass=\Tr\brac{\Sigma}\,.
\end{equation}

The matrix
\noindent
\begin{equation}
\label{eq:66}
\sqrt{\sigma_y\otimes\sigma_y}=\frac{1}{2}
\brac{\begin{array}{cccc}
1+i & 0 & 0 & -1+i\\
0 & 1+i & 1-i & 0\\
0 & 1-i & 1+i & 0\\
-1+i & 0 & 0 & 1+i
\end{array}}
\end{equation}
is symmetric, and so is $\sqrt{\Sigma}$, so Eq.~(\ref{eq:64}) is of the form $A^TA=B^TB$, where $A$ and $B$ are nonsingular. This determines uniquely a complex orthogonal matrix $\Omega$, ($\Omega=(BA^{-1})^T$), satisfying $A=\Omega B$, or
\noindent
\begin{equation}
\label{eq:65}
\sqrt{\sigma_y\otimes\sigma_y}XU=\Omega\sqrt{\Sigma}\,.
\end{equation}
Putting together this result and the decomposition of $TU$ of Eq.~(\ref{eq:63}), we end up with the following useful decomposition of the state,
\noindent
\begin{equation}
\label{eq:67}
X=\sqrt{\sigma_y\otimes\sigma_y}^{\dagger}\Omega\sqrt{\Sigma}P_1FP_{2}^{T}T\,.
\end{equation}
Here we have defined the diagonal unitary matrix $F=O_{7}^{T}E^{\dagger}O_7$, and the orthogonal matrices $P_1=O_4O_7$ and $P_2=O_3O_7$. $O_7$ is an orthogonal matrix permuting the phases of $E$ to ensure that $P_2$ (and $P_1$) have determinant one, and that, for the states where the phases of $F$ are as in Eq.~(\ref{eq:62}), the ordering is either as in that equation or with the first and the fourth phase interchanged (only one of these orderings will be consistent with $\det P_2=\det P_1=1$).

In this decomposition $\Sigma$ gives the concurrence of assistance as in Eq.~(\ref{eq:12}), and $F$ tells us whether it is possible to obtain this by local von Neumann measurements, according to the criterion we found, which was that its diagonal entries have to be the phase factors of Eq.~(\ref{eq:62}) for some $\Phi$. 

The product of the last four terms in this decomposition is $U^{\dagger}=P_1FP_{2}^{T}T$, which gives us information about which basis we should use in the optimal local measurement. Applying the mathematical result stated after Eq.~(\ref{eq:57}) to the orthogonal matrix $P_2=TUP_1F$, we get that there exist $U_1,U_2\in SU(2)$ such that
\noindent
\begin{equation}
\label{eq:13}
T^{\dagger}P_2T=U_1\otimes U_2\,.
\end{equation}
But this is also equal to $T^{\dagger}P_2T=UP_1FT$, and for the states where local measurements are indeed found to be sufficient, the diagonal entries of $F$ are now either as in Eq.~(\ref{eq:62}) or as in this equation but with the first and the fourth phase factor interchanged. In both cases we check explicitly that the matrix $FT$ is of the form of a real orthogonal matrix times a diagonal unitary matrix. Hence, according to Eq.~(\ref{eq:37}), the matrix $T^{\dagger}P_2T=UP_1FT$ of Eq.~(\ref{eq:13}) gives a measurement basis with which we can achieve the concurrence of assistance, and, most importantly, this measurement basis corresponds to a local measurement.

Thus, in summary, writing the state of four qubits in the decomposition (\ref{eq:67}) provides a fast way to find its concurrence of assistance ($\Sigma$), to decide whether local projection measurements by C and D are sufficient to achieve this ($F$), and what those local measurements would be ($P_2$).

It is interesting to count the (continuous) degrees of freedom we have in choosing states for which local von Neumann measurements are sufficient, to see how much space these states take up in the full state space. Normalization of the state $X$ implies that $\Tr\brac{XX^{\dagger}}=\Tr\brac{\Omega\Sigma \Omega^{\dagger}}=1$, so the complex orthogonal matrix $\Omega$ has $12-1=11$ degrees of freedom. The singular value matrix, $\sqrt{\Sigma}$, has $4$, the two real orthogonal matrices, $P_1$ and $P_2$ have each $6$. The unimportance of the overall phase led us to assume $\det(F)=1$, so that in general $F$ has $3$ degrees of freedom, thus counting up to the total number of $30=32-2$ degrees of freedom for an arbitrary pure state of four qubits $\ket{\psi}\in{\mathbb{C}}^{16}$. For states where local von Neumann measurements are sufficient, $F$ has only one degree of freedom, namely the $\Phi$ of Eq.~(\ref{eq:62}), giving a total of $28$ degrees of freedom in choosing these states. Thus, although these states span a set of measure zero on the full state space, they have only two continuous degrees of freedom less than the full state space.

\subsection{The $\rank\brac{\Sigma}<4$-cases}
\label{sec:rankbr-cases}

\noindent The above derivation was completely general, except for one point, namely that in Eq.~(\ref{eq:54}) we assumed the form of $U$ of Eq.~(\ref{eq:37}), thus assuming all the singular values of the matrix $Q$ (\ref{eq:43}) to be nonzero, \emph{i.e.} $\rank(\Sigma)=4$. In the $\rank\brac{\Sigma}<4$-cases it is still possible to write a decomposition of $X$ of the form (\ref{eq:67}), but in these cases we have even more freedom in choosing $P_1FP_{2}^{T}$ (and also in choosing $\Omega$, though this freedom is not important). We now consider the different ranks of $\Sigma$ seperately.

\setcounter{subsubsectionc}{-1}
\subsubsection{$\rank\brac{\Sigma}=0$}
\noindent The case $\rank\brac{\Sigma}=0$ is trivial, since in this case $\Cass=0$ and hence also $\Clocass=\ClocassPOVM=0$.

\subsubsection{$\rank\brac{\Sigma}=1$}
\noindent For states where $\rank\brac{\Sigma}=1$ we write $\Sigma={\rm diag}\brac{\sigma,0,0,0}$, so we have $\Cass=\sigma$. From Eq.~(\ref{eq:43}) we have $Q=U^{*}\Sigma U^{\dagger}$ and hence $Q_{ij}=\sigma U_{i1}^{*}U_{j1}^{*}$. According to Eq.~(\ref{eq:40}) a measurement in the standard basis will yield an average concurrence of 
\noindent
\begin{equation}
\label{eq:27}
\bar{{\cal C}}=\sum_{k}\abs{Q_{kk}}=\sigma\sum_{k}\abs{U_{k1}^{*}}^{2}=\sigma\,,
\end{equation}
so we have $\Clocass=\Cass$ for these states. We notice that in fact every measurement will do for these states. Since the standard basis works for all states, all measurement bases will work for any particular state.

\subsubsection{$\rank\brac{\Sigma}=2$}
\noindent For the $\rank\brac{\Sigma}=2$ case we apply the previously mentioned result by Walgate \emph{et al.} \cite{art402}. Writing $\Sigma={\rm diag}\brac{\sigma_1,\sigma_2,0,0}=U^{T}QU$ as before we now have $\Cass=\sigma_1+\sigma_2$. As discussed in Sec.~\ref{sec:conc-assist}, the column vectors of the unitary matrix $U^{*}$ give an optimal joint measurement basis. According to Walgate \emph{et al.} we can write the first two (orthogonal) column vectors in the form
\noindent
\begin{align}
U_{i1}^{*}&=c_1\ket{0}_C\ket{\eta_0}_D+c_2\ket{1}_C\ket{\eta_1}_D\,,\nonumber\\
U_{i2}^{*}&=c_3\ket{0}_C\ket{\eta_{0}^{\perp}}_D+c_4\ket{1}_C\ket{\eta_{1}^{\perp}}_D\,,\label{eq:28}
\end{align}
with $\braket{\eta_0}{\eta_{0}^{\perp}}=\braket{\eta_1}{\eta_{1}^{\perp}}=0$. A local measurement by C in the standard basis followed by a local measurement by D in either the $\{\ket{\eta_{0}},\ket{\eta_{0}^{\perp}}\}$ or the $\{\ket{\eta_{1}},\ket{\eta_{1}^{\perp}}\}$ basis is equivalent to a joint measurement in the basis 
\noindent
\begin{equation}
\label{eq:9}
\{\ket{0}_C\ket{\eta_0}_D,
\ket{1}_C\ket{\eta_1}_D,
\ket{0}_C\ket{\eta_{0}^{\perp}}_D,
\ket{1}_C\ket{\eta_{1}^{\perp}}_D\}\,.
\end{equation}
The unitary matrix $V$, defined such that the column vectors of $V^{*}$ are these four vectors, is indeed of the form (\ref{eq:44}), and hence the average concurrence after this measurement is
\noindent
\begin{equation}
\label{eq:29}
\bar{{\cal C}}=\sum_{k}\abs{\brac{V^{T}QV}_{kk}}=
\sum_{k}\abs{\brac{W^{*}\Sigma W^{\dagger}}_{kk}}\,,
\end{equation}
where we have defined the matrix $W=V^{\dagger}U$. From Eq.~(\ref{eq:28}) we see that the first two column vectors of $W^{*}$ are $W_{i1}^{*}=\brac{c_1,c_2,0,0}^{T}$ and $W_{i2}^{*}=\brac{0,0,c_3,c_4}^{T}$, so the average concurrence after this measurement is
\noindent
\begin{equation}
\label{eq:30}
\bar{{\cal C}}=\sum_{k}\abs{\sigma_1(W_{k1}^{*})^2+\sigma_2(W_{k2}^{*})^2}
=\sigma_1\brac{|c_1|^2+|c_2|^2}+\sigma_2\brac{|c_3|^2+|c_4|^2}
=\sigma_1+\sigma_2\,,
\end{equation}
attaining the upper bound $\Cass$, and hence $\Clocass=\Cass$ also for these states. Even though classical communication between the assistant parties is needed in this scheme, we know from Sec.~\ref{sec:class-comm-not} that it is possible to find another local measurement basis where classical communication is not necessary.

\subsubsection{$\rank\brac{\Sigma}=3$}
\noindent For $\rank(\Sigma)=3$ the extra freedom in the choice of $U$ of Eq.~(\ref{eq:54}) is only
\noindent
\begin{equation}
\label{eq:60}
U\longrightarrow U
\brac{\begin{array}{cccc}
1 & 0 & 0 & 0\\
0 & 1 & 0 & 0\\
0 & 0 & 1 & 0\\
0 & 0 & 0 & e^{i\varphi}
\end{array}}\,,
\end{equation}
but since we require $\det(U)=1$, we have no more freedom in this case than in the general case, and hence the results for the $\rank(\Sigma)=4$ states apply here as well. So here we have $\Clocass<\Cass$ in general.

\section{Numerical Results}
\label{sec:numerical-results}
\noindent In the previous section we have identified the set of states for which the concurrence of assistance can be achieved by local von Neumann measurements. We found that even though there was quite a lot of freedom in choosing these states, the set is of measure zero on the full state space, so for a randomly chosen state $\Cass>\Clocass$.

It would be interesting to see more quantitatively how big the advantage of joint measurements is. So far we have seen that there exist examples for all the extreme cases $\Clocass=\Cass=1$, $\Clocass=\Cass=0$, and $\Clocass=0, \Cass=1$. But how large can we expect the advantage of joint measurements to be for an arbitrary state? 

To answer this question we have calculated the two quantities, $\Cass$ and $\Clocass$, for a large number ($\sim 10^6$) of arbitrary states. Our earlier analysis provides us with a numerical method to find $\Clocass$ for a given state, Eq.~(\ref{eq:79}), and the concurrence of assistance is easily found from the formula (\ref{eq:5}). Picking pure states of four qubits arbitrarily is done by choosing the individual components $c_k$, Eq.~(\ref{eq:33}), as independent random gaussian variables with mean value zero, each multiplied with a uniformly distributed phase factor \cite{MacKeown}. The joint distribution function of these is indeed spherically symmetric, and therefore, when we have normalized the states, they will give a uniform distribution on the unit sphere in ${\mathbb{C}}^{16}$, which is what we are looking for.

\begin{figure} [htbp]
\vspace*{13pt}
\centerline{\psfig{file=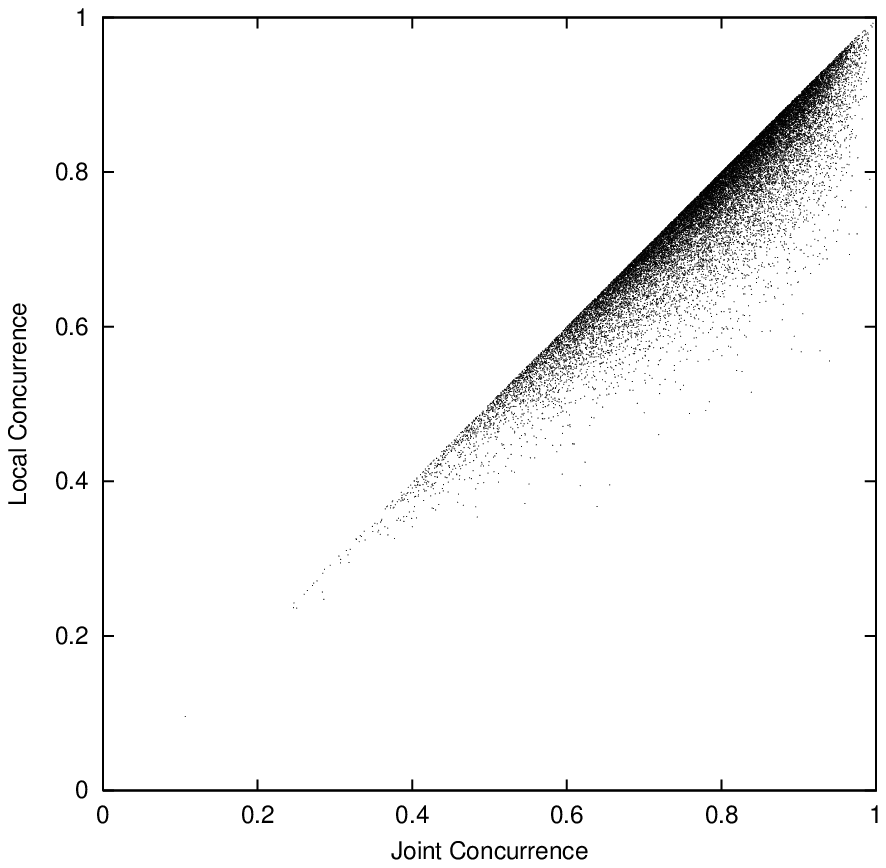, width=8.2cm}} %100 percent
\vspace*{13pt}
\fcaption{\label{fig:stat3}
The distribution of states according to $\Cass$ and $\Clocass$, ($25000$ trial states).}
\end{figure}
\begin{figure} [htbp]
\vspace*{13pt}
\centerline{\psfig{file=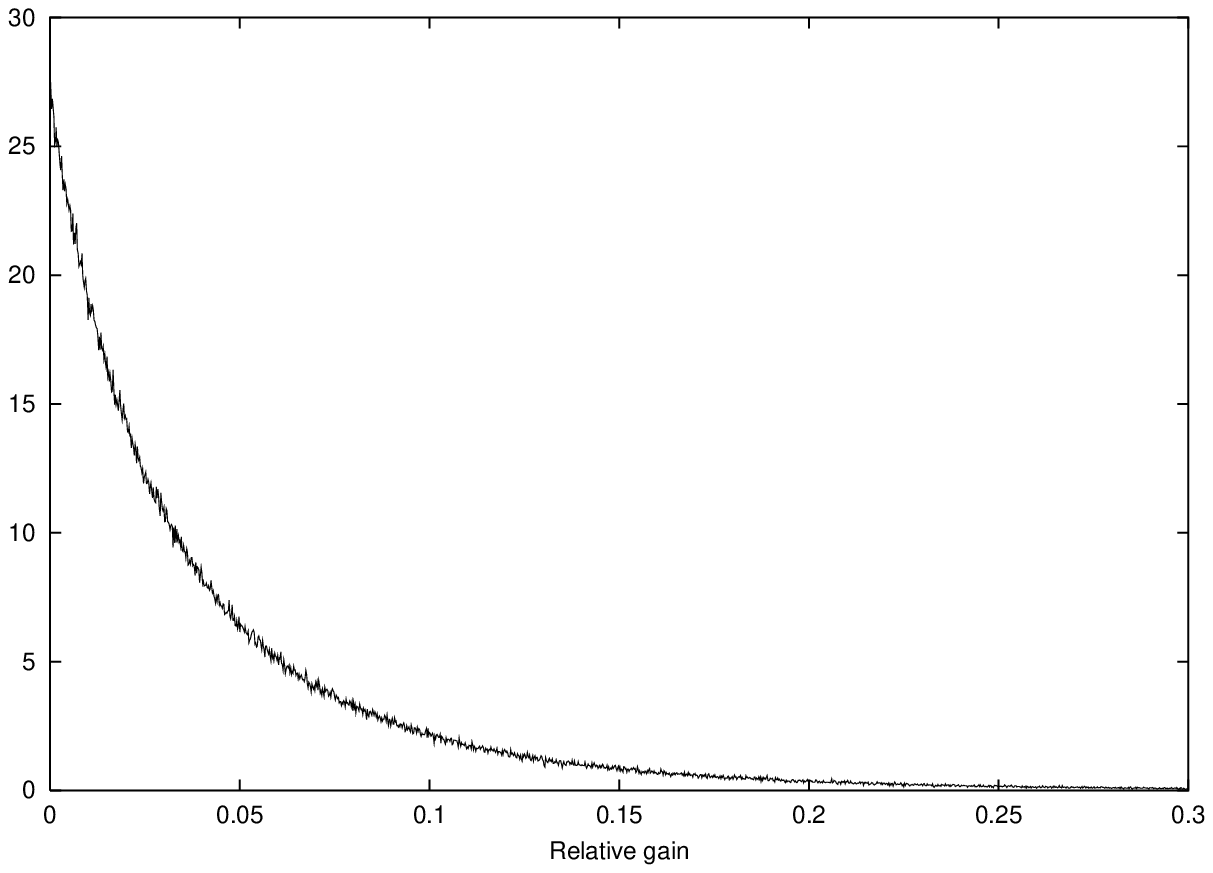, width=8.2cm}} %100 percent
\vspace*{13pt}
\fcaption{\label{fig:diff}
The distribution function for the relative gain, $(\Cass-\Clocass)/\Clocass$, ($10^6$ trial states).}
\end{figure}
\begin{figure} [htbp]
\vspace*{13pt}
\centerline{\psfig{file=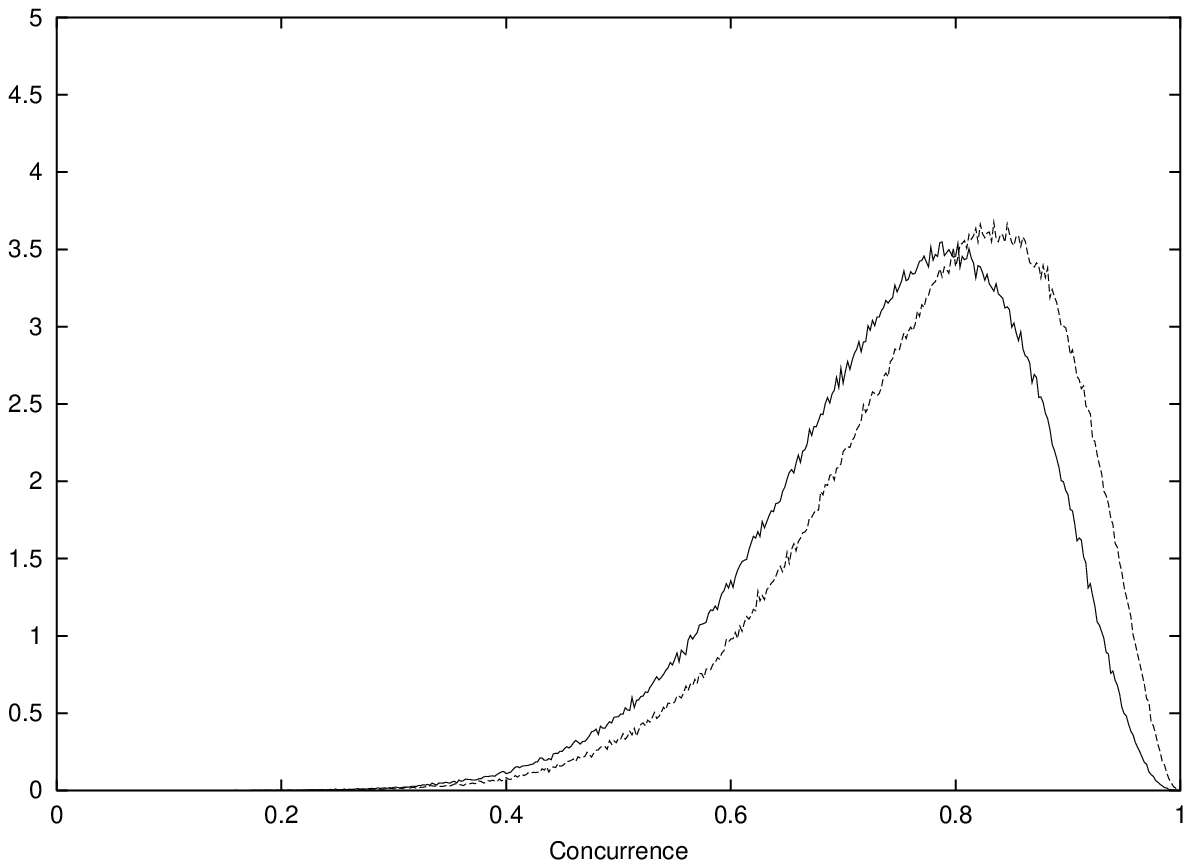, width=8.2cm}} %100 percent
\vspace*{13pt}
\fcaption{\label{fig:stat2}
The distribution functions for $\Cass$ (solid) and $\Clocass$ (dashed), ($10^6$ trial states).}
\end{figure}

The results are presented in Fig.~\ref{fig:stat3}, showing the distribution of $25000$ arbitrarily chosen states plotted according to their values of $\Cass$ and $\Clocass$ (plotting all $10^6$ states doesn't change the appearance of this figure). Here we see a clear accumulation of points close to the limiting $\Cass=\Clocass$ line, indicating that the gain of using joint measurements is in fact typically quite small. This is further substantiated by figure~\ref{fig:diff}, showing the distribution function $P(x)$ for the relative gain, $x=(\Cass-\Clocass)/\Clocass$, normalized so that $\int P(x)dx=1$. Here we see that $P(x)$ is decreasing rapidly (roughly exponentially) for larger $x$ and the average gain is calculated to be only
\noindent
\begin{equation}
\label{eq:14}
\mean{\frac{\Cass-\Clocass}{\Clocass}}=4.6\%\,.
\end{equation}

For a comparison of the two functions $\Cass$ and $\Clocass$, we have also plotted their (normalized) distribution functions separately, Fig.~\ref{fig:stat2}. The mean values of these distributions are found to be
\noindent
\begin{equation}
\label{eq:15}
\langle\Cass\rangle=0.778\hspace{.5cm}{\rm and}\hspace{.5cm}
\langle\Clocass\rangle=0.745\,.
\end{equation}

\begin{figure} [htbp]
\vspace*{13pt}
\centerline{\psfig{file=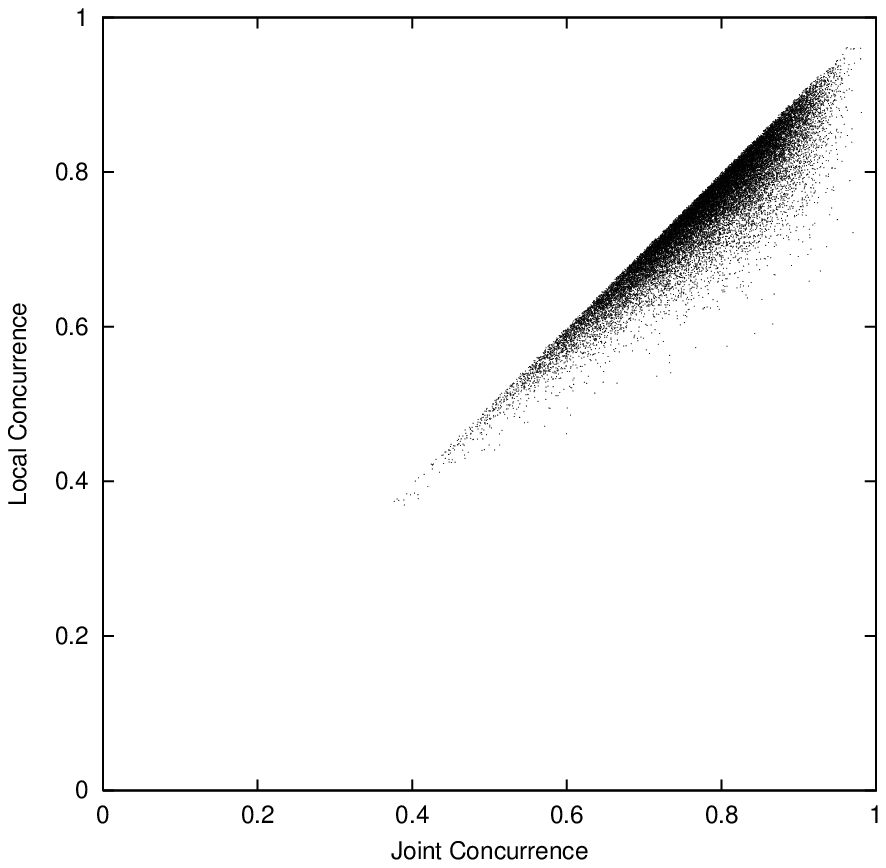, width=8.2cm}} %100 percent
\vspace*{13pt}
\fcaption{\label{fig:stat3_6p}
The distribution of states according to the average of $\Cass$ and $\Clocass$ over the six possible pairs of keepers of the final state, ($25000$ trial states).}
\end{figure}
\begin{figure} [htbp]
\vspace*{13pt}
\centerline{\psfig{file=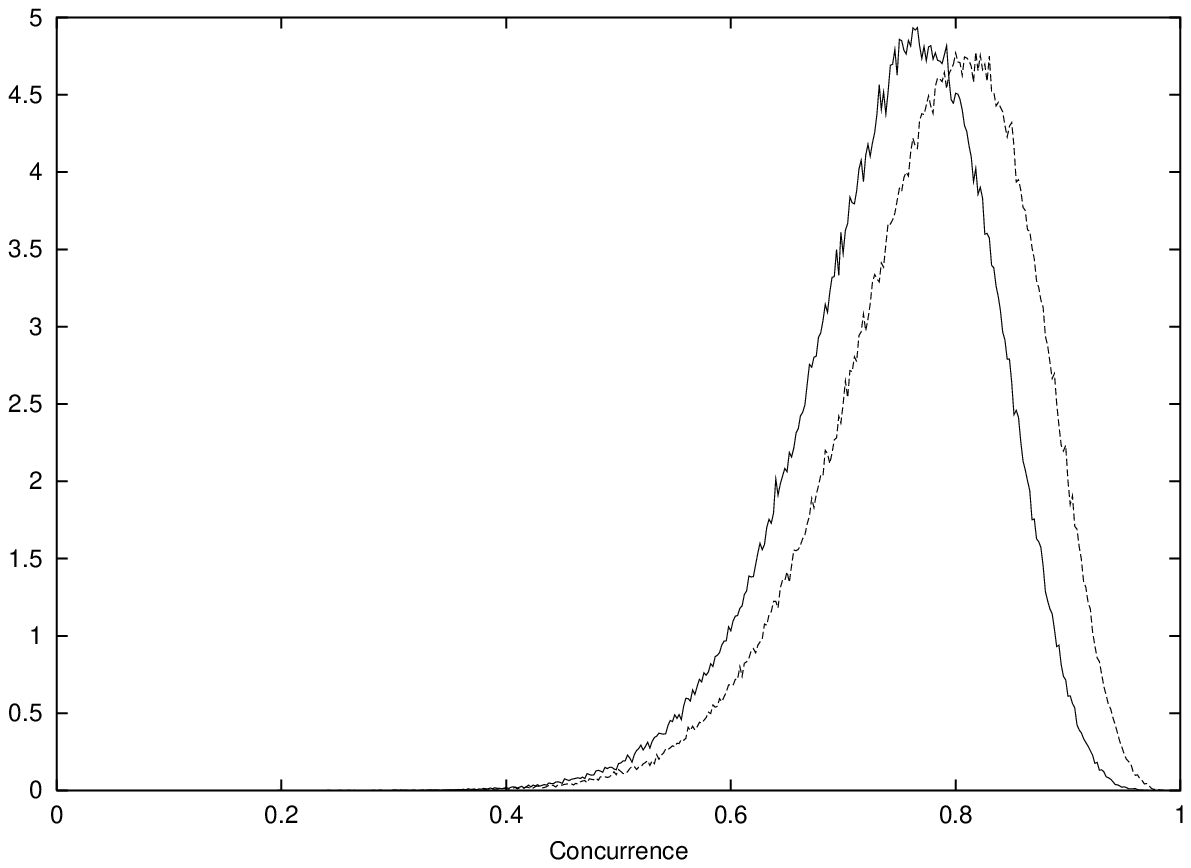, width=8.2cm}} %100 percent
\vspace*{13pt}
\fcaption{\label{fig:stat2_6p}
The distribution functions for the average of $\Cass$ (solid) and $\Clocass$ (dashed) over the six possible pairs of keepers of the final state, ($10^6$ trial states).}
\end{figure}

So far the problem we have considered has been for the assistant parties, C and D, to make A and B end up with as much entanglement (concurrence) as possible. Another interesting quantity would be the average of $\Cass$ and $\Clocass$ over the six possible pairs of keepers of the final state. The figures~\ref{fig:stat3_6p},~\ref{fig:stat2_6p} show the distributions of these quantities. The mean values of the two distributions in Fig.~\ref{fig:stat2_6p} will necessarily be the same as those in Eq.~(\ref{eq:15}) since the states are chosen at random, but we notice that the variances are considerably smaller. This indicates that states with high $\Cass$ or $\Clocass$ for one pair tend to have low $\Cass$ or $\Clocass$ for some of the other pairs and \emph{vice versa}. We stress that these are the results for \emph{arbitrary} states. However, we can find examples of states violating the general tendencies. One would be the state $\ket{\psi}=\ket{0000}+\ket{1111}$, for which local measurements by any two parties can produce a maximally entangled state shared by the other two, and another would be a separable state like $\ket{\psi}=\ket{0000}$, for which it is not possible for any parties to produce any entanglement at all.

\subsection{Nonprojective measurements}
\label{sec:nonpr-meas}
\noindent An interesting extension of the current work would be to investigate the effect of \emph{nonprojective} POVMs. If we allow joint measurements we have seen that von Neumann measurements are sufficient, see the discussion following Eq.~(\ref{eq:36}), but are these also sufficient to achieve the concurrence of local assistance, $\ClocassPOVM$?

At most one of the assistant parties (the one who measures first) needs to do a nonprojective measurement. For after the first measurement the total state is a pure $3$-party state and the problem is reduced to that of finding the concurrence of assistance for this state. For this we know from Sec.~\ref{sec:conc-assist} that a von Neumann measurement is sufficient. 

We have searched among states and measurement schemes where the first measuring party (C) applies a four-outcome POVM, and we have indeed found examples of states where a nonprojective measurement would yield more average concurrence than the best von Neumann measurement on the same state, \emph{i.e.} $\ClocassPOVM>\Clocass$. A particularly simple example is the state
\noindent
\begin{equation}
\label{eq:31}
\ket{\psi}_{{\rm ABCD}}=\frac{1}{\sqrt{3}}
   \Big(\ket{\Phi^{+}}_{{\rm AB}}\ket{++}_{{\rm CD}}
       +\ket{\Phi^{-}}_{{\rm AB}}\ket{00}_{{\rm CD}}
       +\ket{\Psi^{-}}_{{\rm AB}}\ket{11}_{{\rm CD}}\Big)\,,
\end{equation}
for which $\Clocass=0.8801$ whereas $\ClocassPOVM\ge 0.8978$, \emph{i.e.} an increase of about $2.0\%$ (for this state $\Cass=0.9205$). The inequality sign stems from the fact that we have only been considering four-outcome POVMs. The largest difference we have found is for a randomly generated state, for which we saw an increase of $3.5\%$. For random states the increase is typically several orders of magnitude smaller than this. For certain states we have found that the maximum average concurrence is different if D measures first, indicating that classical communication between the two assistant parties might be necessary in doing the optimal nonprojective measurement.

\section{Discussion and Open Questions}
\label{sec:discussion}
\noindent We have studied the concurrence of assistance for $4$-qubit states where $2$ parties are assisting. We introduced three versions of the concurrence of assistance, namely $\Cass$ where the assistants are allowed to perform joint measurements, $\ClocassPOVM$ where they are restricted to local measurements, and $\Clocass$ where they are restricted to local von Neumann measurements. 

For joint measurements we have obtained the simple formula Eq.~(\ref{eq:36}) for the concurrence of assistance, $\Cass$. For local von Neumann measurements we have found that classical communication of measurement outcomes between the two assistant parties is not necessary, but for four-outcome POVMs we have seen that communication does help. A main result is the decomposition of Eq.~(\ref{eq:67}), telling us (i) the concurrence of assistance, (ii) whether there is a local von Neumann measurement that can be used to achieve this and (iii) what that local measurement would be. By using numerical methods we have seen that even though we can construct states for which the gain of concurrence in using joint measurements is anywhere between $0$ and $1$, for the typical state $\Cass$ is only around $5\%$ higher than $\Clocass$. Finally we have seen that in general nonprojective measurements can do a little better than von Neumann measurements, \emph{i.e.} the inequality $\ClocassPOVM>\Clocass$ is strict.

Several open questions arise naturally from the present work. The criterion of Sec.~\ref{sec:what-states-are} to determine the set of states for which local measurements are sufficient to achieve the concurrence of assistance only takes into account von Neumann measurements. Whether there are states for which $\Clocass<\ClocassPOVM=\Cass$, and whether there are states for which $\Cass>\ClocassPOVM=\Clocass$ are both open questions. Other interesting problems would be to see whether communication between the assistant parties is needed for some of the states where $\Clocass<\ClocassPOVM$. Finally, an obvious question is what the results are if we use the entanglement (\ref{eq:1}) as our measure instead of the concurrence.

\nonumsection{Acknowledgments}
\noindent T.L. acknowledges support by The Danish Research Foundation -- Danmarks
Grundforskningsfond and F.V. acknowledges support by DIMACS, the National
Science Foundation's Center for Discrete Mathematics and Theoretical
Computer Science, grant number NSF EIA 00-80234. We thank Chris Fuchs and
Terry Rudolph for useful discussions.

\nonumsection{References}

\appendix{: Proof of Eq.~(\ref{eq:10})}
\label{sec:proof-eq.-refeq:10}
\noindent This is a generalization of a result in Ref.~\cite{HornJohnson} (p.~430).

Let $Q$ be a square matrix with singular values $\sigma_k$, and write its Takagi decomposition (\ref{eq:43}) as $U^{T}QU=\Sigma$. Let $V$ represent an arbitrary right unitary matrix, $VV^{\dagger}=\id$. Then
\noindent
\begin{align}
\max_{V}\abs{\Tr\brac{V^{T}QV}}
&=\max_{V}\abs{\Tr\brac{(U^{\dagger}V)^{T}\Sigma (U^{\dagger}V)}}\nonumber\\
&=\max_{V}\abs{\Tr\brac{V^{T}\Sigma V}}
=\max_{V}\abs{\Tr\brac{\Sigma VV^{T}}}\nonumber\\
&=\max_{V}\abs{\sum_{k}\sigma_k\brac{VV^{T}}_{kk}}
\le\max_{V}\sum_{k}\sigma_k\abs{\brac{VV^{T}}_{kk}}
\le\sum_{k}\sigma_k\,,\label{eq:16}
\end{align}
where the last inequality follows from the fact that
\noindent
\begin{equation}
\label{eq:17}
\abs{\brac{VV^{T}}_{kk}}=\abs{\sum_j \brac{V_{kj}}^2}
\le\sum_j |V_{kj}|^2=(VV^{\dagger})_{kk}=1\,.
\end{equation}

Now assume that there exist a right unitary matrix $V_0$ satisfying
\noindent
\begin{equation}
\label{eq:18}
\sum_k\abs{\brac{V_{0}^{T}QV_0}_{kk}}>\sum_k\sigma_k\,.
\end{equation}
Let its diagonal entries be $(V_{0}^{T}QV_{0})_{kk}=r_{k}e^{i\delta_k}$ with $r_k$ real and positive, and define the matrix $V_1=V_0D$, where $D$ is diagonal and $D_{kk}=e^{-i\delta_k/2}$. Since $V_{1}$ is right unitary, and
\noindent
\begin{equation}
\label{eq:19}
\abs{\Tr\brac{V_{1}^{T}QV_{1}}}=\sum_k r_{k}>\sum_k\sigma_k\,,
\end{equation}
in contradiction with Eq.~(\ref{eq:16}), we conclude that
\noindent
\begin{equation}
\label{eq:20}
\sum_k\abs{\brac{V^{T}QV}_{kk}}\le\sum_k\sigma_k\,,
\end{equation}
for all right unitary matrices $V$.

\appendix{: Proof of Eq.~(\ref{eq:53})}
\label{sec:proof-eq.-refeq:53}
\noindent Let $U$ be unitary. Write the real and imaginary parts of $U$ and find the singular value decompositions of the real part,
\noindent
\begin{equation}
\label{eq:58}
U=U_R+iU_I=\tilde{O}_1\Sigma_{R}\tilde{O}_{2}^{T}+iU_I\,,
\end{equation}
where $\tilde{O}_1$ and $\tilde{O}_2$ are both real and orthogonal. Now look at the matrix $\tilde{U}=\tilde{O}_{1}^{T}U\tilde{O}_{2}^{T}=\Sigma_{R}+iA$. Unitarity, $\tilde{U}\tilde{U}^{\dagger}=\tilde{U}^{\dagger}\tilde{U}=\id$ implies
\noindent
\begin{equation}
\label{eq:23}
\Sigma_{R}^{2}+AA^{T}=\Sigma_{R}^{2}+A^{T}A=\id\hspace{.5cm}{\rm and}\hspace{.5cm}
A\Sigma_R-\Sigma_RA^{T}=0\,,
\end{equation}
so the real matrix $A$ is \emph{normal}, $AA^{T}=A^{T}A$, and hence it has an eigenvalue decomposition \cite{HornJohnson}, 
\noindent
\begin{equation}
\label{eq:22}
A=V\Lambda V^{\dagger}\,,
\end{equation}
with $V$ unitary. The matrix
\noindent
\begin{equation}
\label{eq:24}
AA^{T}=AA^{\dagger}=V\Lambda V^{\dagger}V\Lambda^{*}V^{\dagger}
=V\abs{\Lambda}^2V^{\dagger} 
\end{equation}
is diagonal according to Eq.~(\ref{eq:23}). If all its entries are distinct, $V$ can only be making permutations of these, and hence it can also only do so in Eq.~(\ref{eq:22}), meaning that $A$ is diagonal.

If some of the diagonal entries of $AA^{T}$ are identical, we first apply the real and orthogonal operator $P$ permuting the diagonal entries, to ensure that the identical ones appear last on the diagonal of $AA^{T}$. Thus we define $\tilde{\tilde{U}}=P\tilde{U}P^{T}=\tilde{\Sigma}_{R}+i\tilde{A}$. 

For the special case of two identical and two distinct entries, the matrices are now
\noindent
\begin{align}
\tilde{A}\tilde{A}^{T}&={\rm diag}\brac{\abs{\lambda_1}^2,\abs{\lambda_2}^2,\abs{\lambda_3}^2,\abs{\lambda_3}^2}\,,\nonumber\\
\tilde{\Sigma}_{R}^{2}&={\rm diag}\brac{1-\abs{\lambda_1}^2,1-\abs{\lambda_2}^2,1-\abs{\lambda_3}^2,1-\abs{\lambda_3}^2}\,,\nonumber\\
\tilde{A}&={\rm diag}\brac{\lambda_1,\lambda_2}\oplus A'\,.\label{eq:25}
\end{align}
From the unitarity condition (\ref{eq:23}), $\tilde{A}\tilde{\Sigma}_{R}=\tilde{\Sigma}_{R}\tilde{A}^{T}$ or $\tilde{A}_{ij}\tilde{\sigma}_j=\tilde{\sigma}_i\tilde{A}_{ji}$, we see that the submatrix $A'$ of Eq.~(\ref{eq:25}) is symmetric, so we can write its eigenvalue decomposition $A'=V'\Lambda'{V'}^{T}$ with $V'$ real and orthogonal and with $\Lambda'={\rm diag}\brac{\lambda_3',\lambda_4'}$. Defining the matrix $\tilde{V}=\id\oplus V'$ we finally get
\noindent
\begin{equation}
\label{eq:26}
\tilde{V}P\tilde{O}_1U\tilde{O}_{2}^{T}P^{T}\tilde{V}^{T}=\tilde{V}\tilde{\tilde{U}}\tilde{V}^{T}=D\,,
\end{equation}
where $D=\tilde{\Sigma}_{R}+i{\rm diag}\brac{\lambda_1,\lambda_2,\lambda_3',\lambda_4'}$. 

This generalizes straightforwardly to other degeneracies, and hence we have the desired decomposition (\ref{eq:53}), where we identify $O_1=\tilde{V}P\tilde{O}_1$ and $O_2=\tilde{V}P\tilde{O}_2$.

This decomposition is not completely unique. The eigenvalues and singular values that we have found are unique, so the matrix $D$ is unique up to certain permutations between the real values and between the imaginary values of its entries. For $D$ to remain unitary all the diagonal entries must still be of the form $e^{i\varphi}$, and hence the only allowed form of permutations can be $\Ima{D_{ii}}\leftrightarrow \Ima{D_{jj}}$ with $\Ima{D_{jj}}=-\Ima{D_{ii}}$. Since $\Rea{D_{ii}}>0$ per construction, this is the same as interchanging $D_{ii}\leftrightarrow D_{jj}$, and hence we conclude that $D$ is uniquely defined up to permutations of the diagonal entries, which we shall refer to as the \emph{phases} of $U$, $e^{i\delta_i}$, defined such that $-\frac{\pi}{2}<\delta_i\le\frac{\pi}{2}$. 

For a given ordering of the phases in $D$, the $O_1$ and $O_2$ are unique up to simultaneous transformations of the kind $O_1\to O_1\tilde{O}$, $O_2\to O_2\tilde{O}$, where $\tilde{O}$ is a real, orthogonal matrix leaving $D$ invariant, $\tilde{O}D\tilde{O}^{T}=D$. In the case where all the phases are different, $\tilde{O}$ is a diagonal matrix with $\pm 1$ on the diagonal. Otherwise it is a block diagonal matrix, with the sizes of the blocks determined by the multiplicity of the phases.


\begin{thebibliography}{000}
\bibitem{art211}
D.~P. DiVincenzo, C.~A. Fuchs, H. Mabuchi, J.~A. Smolin, A. Thapliyal, 
  and A. Uhlmann (1999), {\em Entanglement of assistance},
  in {\em Quantum Computing and Quantum Communications}, 
  Vol.~1509 of {\em Lecture Notes in Computer Science}, edited by 
  C.~P. Williams, Springer-Verlag (Berlin), pp.~247--257, 
  ({\tt quant-ph/9803033}).

\bibitem{art292}
O. Cohen (1998), {\em Unlocking hidden entanglement with classical information}, Phys. Rev. Lett. {\bf 80},  pp.~2493--2496.

\bibitem{art129}
L.~P. Hughston, R. Jozsa, and W.~K. Wootters (1993),
 {\em A complete classification of quantum ensembles having a given density matrix}, Phys. Lett. A {\bf 183}, pp.~14--18.

\bibitem{art402}
J. Walgate, A.~J. Short, L. Hardy, and V. Vedral (2000), 
{\em Local distinguishability of multipartite orthogonal quantum states}, 
 Phys. Rev. Lett. {\bf 85}, pp.~4972--4975, see also J. Walgate and L. Hardy (2002), {\em Nonlocality, Asymmetry, and Distinguishing Bipartite States}, {\tt quant-ph/0202034}.

\bibitem{art280}
S. Hill and W.~K. Wootters (1997), {\em Entanglement of a pair of quantum bits}, Phys. Rev. Lett. {\bf 78},  pp.~5022--5025.

\bibitem{art83}
W.~K. Wootters (1998), {\em Entanglement of formation of an arbitrary state of two qubits}, Phys. Rev. Lett. {\bf 80},  pp.~2245--2248.

\bibitem{art82}
G. Vidal (2000), {\em Entanglement monotones}, J. Mod. Opt. {\bf 47},  pp.~355--376.

\bibitem{art148} V. Coffman, J. Kundu, and W. K. Wootters (2000), {\em Distributed entanglement}, Phys. Rev. A {\bf 61}, no.~052306.

\bibitem{art293} K. M. O'Connor and W. K. Wootters (2001), {\em Entangled rings}, Phys. Rev. A {\bf 63}, no.~052302.

\bibitem{art41}
C.~H. Bennett, D.~P. DiVincenzo, J.~A. Smolin, and W.~K. Wootters (1996), 
{\em Mixed-state entanglement end quantum error correction}, 
 Phys. Rev. A {\bf 54},  pp.~3824--3851.

\bibitem{art403}
K. Audenaert, F. Verstraete, and B. {De Moor} (2001), {\em Variational characterizations of separability and entanglement of formation}, Phys. Rev. A {\bf 64},  no.~052304.

\bibitem{NielsenChuang}
M.~A. Nielsen and I.~L. Chuang (2000), {\em Quantum Computation and Quantum
  Information}, Cambridge University Press.

\bibitem{note:chris}
C.~A. Fuchs (1998), unpublished notes.

\bibitem{HornJohnson}
R.~A. Horn and C.~R. Johnson (1985), {\em Matrix Analysis}, Cambridge University Press.

\bibitem{art408}
F. Verstraete, J. Dehaene, B. {De Moor}, and H. Verschelde, Phys. Rev. A {\bf
  65},  052112  (2002).

\bibitem{MacKeown}
P.~K. MacKeown (1997), {\em Stochastic Simulation in Physics}, Springer-Verlag (Singapore).

\end{thebibliography}
\end{document}